\input amstex
\documentstyle{amsppt}
\loadmsbm

\topmatter

\title
Quantum Algorithms for Lowest Weight Paths and Spanning Trees in Complete Graphs
\endtitle

\rightheadtext{Quantum Algorithms for Graphs}

\author
Mark Heiligman
\endauthor

\address
Advanced Research and Development Activity,
Suite 6644,
National Security Agency, 
9800 Savage Road, 
Fort Meade, Maryland 20755
\endaddress

\email
miheili\@nsa.gov
\endemail

\date
January 9, 2003
\enddate

\keywords
quantum computing, quantum algorithm, graph theory
\endkeywords

\abstract
Quantum algorithms for several problems in graph theory are
considered.  Classical algorithms for finding the lowest weight path
between two points in a graph and for finding a minimal weight
spanning tree involve searching over some space.  Modification of
classical algorithms due to Dijkstra and Prim allows quantum search to
replace classical search and leads to more efficient algorithms.  In
the case of highly asymmetric complete bipartite graphs, simply replacing
classical search with quantum search leads to a faster quantum
algorithm.  A fast quantum algorithm for computing the diameter of 
a complete graph is also given.
\endabstract

\endtopmatter

\document

\def\section#1{\heading #1 \endheading}

\def\card#1{\vert#1\,\vert}
\def\bigO#1{O\bigl(#1\bigr)}
\def\littleo#1{o\bigl(#1\bigr)}
\def\R{\Bbb R}
\def\nullset{\{\}}
\def\weight{\nu}

\def\vs{\vskip 1pt}
\def\hs{\hskip 10pt}
\def\lab#1{{\it (#1)}}
\def\getssetto{\leftarrow}

\def\alga{\bf
\lab{1} $S \getssetto \{v_0\}$, $\lambda(v_0) \getssetto 0$
\vs\lab{2} for $v \in V-S$ do
\vs\hs\lab{2.a} $\lambda(v) \getssetto \weight(v_0,v)$
\vs\lab{3} while $S \ne V$ do
\vs\hs\lab{3.a} find $w \in V-S$ such that $\lambda(w)$ is minimal
\vs\hs\lab{3.b} $S \getssetto S \cup \{w\}$
\vs\hs\lab{3.c} for $v \in V-S$ do
\vs\hs\hs\lab{3.c.1} $\lambda(v) \getssetto \min\bigl(\lambda(v),\lambda(w)+\weight(w,v)\bigr)$
}

\def\algb{\bf
\lab{1} $S \getssetto \{v_0\}$, $\lambda(v_0) \getssetto 0$
\vs\lab{2} while $S \ne V$ do
\vs\hs\lab{2.a} find $(w,v) \in S \times (V-S)$ such that $\lambda(w)+\weight(w,v)$ is minimal
\vs\hs\lab{2.b} $S \getssetto S \cup \{v\}$, $\lambda(v) \getssetto \lambda(w)+\weight(w,v)$
}

\def\algc{\bf
\lab{1} $S \getssetto \{v_0\}$, $T \getssetto \{v_0\}$, $\lambda(v_0) \getssetto 0$
\vs\lab{2} for $v \in V-S$ do
\vs\hs\lab{2.a} $\lambda(v) \getssetto \weight(v_0,v)$
\vs\lab{3} while $S \ne V$ do
\vs\hs\lab{3.a} find $(w,v) \in T \times (V-S)$ such that $\lambda(w)+\weight(w,v)$ is minimal
\vs\hs\lab{3.b} find $u\in V-S$ such that $\lambda(u)$ is minimal
\vs\hs\lab{3.c} if $\lambda(w)+\weight(w,v) \le \lambda(u)$ then
\vs\hs\hs\lab{3.c.1} $S \getssetto S \cup \{v\}$, $T \getssetto T \cup \{v\}$, $\lambda(v) \getssetto \lambda(w)+\weight(w,v)$
\vs\hs\lab{3.d} if $\lambda(w)+\weight(w,v) > \lambda(u)$ then
\vs\hs\hs\lab{3.d.1} $S \getssetto S \cup \{u\}$, $T \getssetto T \cup \{u\}$
\vs\hs\lab{3.e} if $\card{T} \ge k$ then do
\vs\hs\hs\lab{3.e.1} for $v \in V-S$ do
\vs\hs\hs\hs\lab{3.e.1.a} find $w\in T$ such that $\lambda(w)+\weight(w,v)$ is minimal
\vs\hs\hs\hs\lab{3.e.1.b} if $\lambda(w)+\weight(w,v) < \lambda(v)$ then
\vs\hs\hs\hs\hs\lab{3.e.1.b.1} $\lambda(v) \getssetto \lambda(w)+\weight(w,v)$
\vs\hs\hs\lab{3.e.2} $T \getssetto \{v_0\}$
}



\def\alge{\bf
\lab{1} $S_2 \getssetto \{v_0\}$, $\lambda_1(v_0) \getssetto 0$
\vs\lab{2} for $v \in V_1-S_1$ do
\vs\hs\lab{2.a} find $w \in V_2$ such that $\weight_1(v_0,w)+\weight_2(w,v)$ is minimal
\vs\hs\lab{2.b} $\lambda_1(v) \getssetto \weight_1(v_0,w)+\weight_2(w,v)$
\vs\lab{3} while $S_1 \ne V_1$ do
\vs\hs\lab{3.a} find $(u,v,w) \in S_1 \times V_2 \times (V_1-S_1)$ such that $\lambda_1(u)+\weight_1(u,v)+\weight_2(v,w)$ \vs\hs\hs\hs\hs is minimal
\vs\hs\lab{3.b} $S_1 \getssetto S_1 \cup \{w\}$, $\lambda_1(w) \getssetto \lambda_1(u)+\weight_1(u,v)+\weight_2(v,w)$
\vs\hs\lab{3.c} for $v \in V_1-S$ do
\vs\hs\hs\lab{3.c.1} find $u \in V_2$ such that $\lambda_1(w)+\weight_1(w,u)+\weight_2(u,v)$ is minimal
\vs\hs\hs\lab{3.c.2} $\lambda_1(v) \getssetto \min\bigl(\lambda_1(v),\lambda_1(w)+\weight_1(w,u)+\weight_2(u,v)\bigr)$
\vs\lab{4} for $u \in V_2$ do
\vs\hs\lab{4.a} find $w \in V_1$ such that $\lambda_1(w)+\weight_1(w,u)$ is minimal
\vs\hs\lab{4.b} $\lambda_2(u) \getssetto \lambda_1(w)+\weight_1(w,u)$
}

\def\algf{\bf
\lab{1} $S \getssetto \{v_0\}$, $F \getssetto \nullset$
\vs\lab{2} for $v \in V-S$ do
\vs\hs\lab{2.a} $L(v) \getssetto \weight(v_0,v)$, $M(v) \getssetto v_0$
\vs\lab{3} while $S \ne V$ do
\vs\hs\lab{3.a} find $w \in V-S$ such that $L(w)$ is minimal
\vs\hs\lab{3.b} $S \getssetto S \cup \{w\}$, $F \getssetto F \cup \{(w,M(w)\}$
\vs\hs\lab{3.c} for $v \in V-S$ do
\vs\hs\hs\lab{3.c.1} if $\weight(w,v) < L(v)$ then
\vs\hs\hs\hs\lab{3.c.1.a} $L(v) \getssetto \weight(w,v)$, $M(v) \getssetto w$
}

\def\algg{\bf
\lab{1} $S \getssetto \{v_0\}$, $F \getssetto \nullset$
\vs\lab{2} while $S \ne V$ do
\vs\hs\lab{2.a} find $(u,v) \in S \times (V-S)$ such that $\weight(u,v)$ is minimal
\vs\hs\lab{2.b} $S \getssetto S \cup \{v\}$, $F \getssetto F \cup \{(u,v)\}$ 
}

\def\algh{\bf
\lab{1} $S \getssetto \{v_0\}$, $T \getssetto \{v_0\}$, $F \getssetto \nullset$, $L(v_0) \getssetto 0$
\vs\lab{2} for $v \in V-S$ do
\vs\hs\lab{2.a} $\lambda(v) \getssetto \weight(v_0,v)$, $M(v) \getssetto v_0$
\vs\lab{3} while $S \ne V$ do
\vs\hs\lab{3.a} find $(u,v) \in T \times (V-S)$ such that $\weight(u,v)$ is minimal
\vs\hs\lab{3.b} find $w \in V-S$ such that $L(w)$ is minimal
\vs\hs\lab{3.c} if $L(w) \le \weight(u,v)$ then
\vs\hs\hs\lab{3.c.1} $S \getssetto S \cup \{w\}$, $T \getssetto T \cup \{w\}$, $F \getssetto F \cup \{(w,M(w)\}$
\vs\hs\lab{3.d} if $L(w) > \weight(u,v)$ then
\vs\hs\hs\lab{3.d.1} $S \getssetto S \cup \{v\}$, $T \getssetto T \cup \{v\}$, $F \getssetto F \cup \{(u,v)\}$
\vs\hs\lab{3.e} if $\card{T} \ge k$ then do
\vs\hs\hs\lab{3.e.1} for $v \in V-S$ do
\vs\hs\hs\hs\lab{3.e.1.a} find $w\in T$ such that $\weight(w,v)$ is minimal
\vs\hs\hs\hs\lab{3.e.1.b} if $\weight(w,v) < L(v)$ then
\vs\hs\hs\hs\hs\lab{3.e.1.b.1} $L(v) \getssetto \weight(w,v)$, $M(v) \getssetto w$
\vs\hs\hs\lab{3.e.2} $T \getssetto \{v_0\}$
}


\head
Introduction
\endhead

The question of which classical algorithms can be sped up by quantum
computing is of course a very interesting one.  At present there are
only a few general techniques known in the field of quantum computing
and finding new problems that are amenable to quantum speedups is a
high priority.  Classically, one area of mathematics that is full of
interesting algorithms is computational graph theory.  It is therefore
natural to ask whether any of the classical graph theory algorithms
can take advantage of quantum computing.


One of the few general techniques known centers around Grover's
algorithm for searching an unsorted list for a specified element.
This original idea has been extended to general amplitude
amplification that can be applied to any classical algorithm.  It
would be incorrect to assume that amplitude amplification always leads
to quantum speedups of classical algorithms.  There are some
interesting cases where ``Grover-like'' techniques do in fact lead to
speedups of classical algorithms.  One very important case
of this is to find the minimum value of a computable function as
the set of input arguments ranges over a finite, but unordered list.
In this case, if the list is of length $n$, then the quantum cost of
finding the minimum is $\bigO{\sqrt{n}}$, while the classical cost is
$\bigO{n}$.  Quantum algorithms for searching for the maximum or minimum
of an unsorted list have been described in \cite{DH} and \cite{AK}.
The question to be addressed here is whether this leads to
speedups in graph theory algorithms that employ classical minimum
finding in the course of solving a graph theory problem.
 
\head
1. Minimal Weight Paths
\endhead

The problem of finding the shortest path between two points in a
weighted graph is an old one.  If $G=(V,E)$ is a directed graph with a
weighting function $\weight: E \rightarrow \R^+$, the weight of a path
is the sum of the weights of the edges that comprise the path.  
If $G$ is a complete graph, then the function $\weight$ is well defined on
all the edges.  In case $G$ is not a complete graph, and $(v,v^\prime)
\notin E$, then it is useful to define $\weight(v,v^\prime) = \infty$.  This 
allows $\weight$ to be defined on all of $V \times V$, not just $E$.

Pick a $v_0\in V$ from which all the shortest paths are to be computed.
Consider the following algorithm due to Dijkstra (see \cite{G}):
$$\vbox{\vskip 10pt
\vbox{\alga}
\vskip 10pt\hskip 30pt
\hbox{\it Figure 1: Dijkstra's algorithm.}
}$$
At the end of this procedure $\lambda(v)$ is the length of the
shortest path from $v_0$ to $v$, and with only minor changes,
this algorithm is easily modified to record the information needed to
construct the shortest path.  This algorithm consists of iterating
over all of $V$ by successively adding elements to $S$, which is the
set of points for which the shortest path from $v_0$ has already been
determined.  Each iteration consists of a search procedure to find the
next nearest element of $V$ to $v_0$, and an update procedure to
record all the newest shortest path information for the remaining
vertices based on this newest next nearest element.

Analysis of the run time for this algorithm is quite simple.  At line
(3) if $i=\card{S}$, then the (classical) cost of the search in line
(3a) is $n-i$ and the cost of updating in line (3c) is $n-i-1$.  The
total cost is therefore $\sum_{i=0}^n 2n-2i-1 = \bigO{n^2}$.

There are some important modifications to this original algorithm of
Dijkstra if the graph being searched is somewhat sparse (i.e. if the
total number of edges is much less than $n^2$).  In this case,
Dijkstra's algorithm can be modified to find the shortest path with
work $\bigO{(\card{V}+\card{E})\log\card{V}}$ through the use of
priority queues.  In general, the key idea is that the update
procedure in line (3c) need only update the shortest path for those
vertices that are adjoining the vertex $w$ which was most recently
added to the set $S$ of those vertices whose shortest path distance
from $v_0$ have alread been computed.  The reason the only the
vertices adjacent to $w$ need to be considered is that for all other
vertices $\weight(w,v)=\infty$.

Unfortunately, if the classical search for the
minimum at line (3a) is replaced by the quantum algorithm for finding
the minimum of an unordered set, the cost for line (3a) per iteration
drops to $\bigO{\sqrt{n-i}}$, but the cost of the entire algorithm is still
$\bigO{n^2}$ since the update cost per iteration in line (3c) is still
$\bigO{n-i}$.

One possible way around this problem is to dispense entirely with the
update procedure in line (3c) at the cost of a larger search in line
(3a).  This modified algorithm then goes as follows:
$$\vbox{\vskip 10pt
\vbox{\algb}
\vskip 10pt\hskip 30pt
\hbox{\it Figure 2: Dijkstra's algorithm without full updating}
}$$
As before, at the end of this procedure $\lambda(v)$ is the length of
the shortest path from $v_0$ to $v$, and again, this algorithm is 
easily modified to record the information needed to construct the shortest path.

The analysis of this modified algorithm is also quite easy.  At line
(2) if $i=\card{S}$, then the (classical) cost of the search in line
(2a) is $i\,(n-i)$, making the entire classical cost of this algorithm
$\sum_{i=0}^n i\,(n-i) = \bigO{n^3}$, which is quite a bit worse than
Dijkstra's original algorithm.  However, the quantum cost of this
algorithm is determined by noting that the search cost at line (2a)
reduced to $\bigO{\sqrt{i\,(n-i)}}$, thereby making the total
cost of the algorithm $\bigO{\sum_{i=0}^n \sqrt{i\,(n-i)}} =
\bigO{n^2}$, which can be seen by noting that
$$\sum_{i=0}^n \sqrt{i\,(n-i)} 
\approx \int_{0}^{n} \sqrt{x\,(n-x)}\, dx
=n^2\,\int_0^1 \sqrt{y\,(1-y)}\, dy = \bigO{n^2}.$$
So the quantum version of this algorithm has work $\bigO{n^2}$ as well,
which really doesn't represent an improvement over the original
classical algorithm.  

What seems to be really needed is to have a partial tradeoff between
the search and update parts of the algorithm.  The idea is to balance
the classical update cost per iteration with the quantum search cost.
The following algorithm is one way of accomplishing this.
$$\vbox{\vskip 10pt
\vbox{\algc}
\vskip 10pt\hskip 30pt
\hbox{\it Figure 3: Dijkstra's algorithm with periodic updating}
}$$
The idea is to have a set $T$ of vertices for which the full update of
$\lambda(v)$ for all remaining vertices in $V-S$ has not yet been
computed.  This full updating is done every $k$-th iteration of the
main loop.  By keeping $v_0$ in $T$ all the time, there is always the
possibility of going directly from $v_0$ to $v$ since $\lambda(v_0)=0$.
The value of $k$ is a parameter for this algorithm and needs to be set
to optimize the total cost.

Most of the work in this algorithm takes place in line (3), the main
iteration, which is done a total of $n$ times.  To analyze the work
for the $i$-th iteration, write $i=h\,k+j$ with $1 \le j \le k$, so
that the size of the set $T$ on the $i$-th iteration is $j$.  It is
convenient for this analysis to assume that $k$ divides $n$ since that
makes line (3c) execute exactly $n/k$ times, however, even if this
assumption does not hold, the work calculation is still valid.

The search for the minimum on line (3a) is over a set of size
$j\,(n-i)$, so the total work over all iteration of line (3a) is
$$\sum_{h=0}^{n/k-1}\sum_{j=1}^k j\,(n-h\,k-j) = \bigO{kn^2}$$
in the classical case and
$$\sum_{h=0}^{n/k-1}\sum_{j=1}^k \sqrt{j\,(n-h\,k-j)} = \bigO{k^{1/2}n^{3/2}}$$
in the quantum case.

The update cost on line (3e1) requires $n-i$ searches for the minimum
over a set of size $k$ each time and this is done only when $i$ is
divisible by $k$.  Therefore the entire cost of the update procedure
on line (3e) over all the iterations of the algorithm is 
$$\sum_{h=1}^{n/k} (n-kh)\,k = \bigO{n^2}$$
in the classical case
and $$\sum_{h=1}^{n/k} (n-kh)\,\sqrt{k} = \bigO{k^{-1/2}n^2}$$
in the quantum case.

In the classical case, the total work for the algorithm is 
$\max\bigl(\bigO{kn^2},\bigO{n^2}\bigr)=\bigO{n^2}$
which is minimized by taking $k=1$, and therefore
$$W_{\hbox{\sevenrm classical}}=\bigO{n^2}.$$ 
Note that taking $k=1$ in figure 3 gives the original Dijkstra
algorithm of figure 1, while taking $k=\card{V}$ yields the algorithm
of figure 2, so $k$ might reasonably be viewed as an interpolation 
parameter.

In the quantum case, the situation is a bit different.  The total work
in line (3) is just the maximum of the work in lines (3a) and (3e), since
the work in line (3b) is always dominated by these other work factors.
The total work is therefore
$$\max\bigl(\bigO{k^{1/2}n^{3/2}},\bigO{k^{-1/2}n^2}\bigr)$$ 
and to minimize this, the parameter $k$ should be chosen to make these
two work factors the same.  Setting $k^{1/2}n^{3/2} = k^{-1/2}n^2$ gives
$k=n^{1/2}$ and therefore
$$W_{\hbox{\sevenrm quantum}}=\bigO{n^{7/4}}.$$
This indeed is an improvement over the classical work factor of $n^2$.

\head
2. Graph Diameter
\endhead

The algorithms so far presented compute the minimal path length from
one point in the graph to another point in the graph.  This minimal
path length will be referred to as the distance from one point to
another in the graph.  The diameter of a graph is the distance between
the two furthest points in a graph (i.e. the maximum of the minimal
path lengths in the graph).  This section deals with quantum algorithms
for finding the diameter of a weighted graph.

Since the initial point $v_0$ was always fixed in Dijkstra's algorithm
and its variants, this was never explicitly indicated in the notation
$\lambda(v)$.  However for the purposed of this section, it is useful
to write $\lambda(v_0,v)$ in place of $\lambda(v)$.  One of the key
features of all of the ``Dijkstra-like'' algorithms described in the
first section is that they start from a given vertex $v_0$ and by an
iterative procedure manage to find the minimal distances to all the
other vertices in the graph.  Furthermore the last vertex for which the
distance is computed is always the most distant vertex from $v_0$.
These ``Dijkstra-like'' algorithms could therefore be viewed as
computing the maximum distance from $v_0$. The diameter of the graph
is just the maximum of all these distances as $v_0$ runs over all of
$V$.  By invoking the quantum maximum finding algorithm with a
``Dijkstra-like'' algorithm as a callable subroutine, the diameter of
the graph will follow.  Since there are $n=\card{V}$ possible initial
values of $v_0$, the quantum cost is simply $\sqrt{n}$ times the cost
of the inner loop.  Using the best quantum algorithm as the inner loop
gives a total quantum cost for finding the diameter as $\bigO{n^{9/4}}$.

By way of comparison with classical costs, Dijkstra's (classical)
algorithm would have to be run $n$ times giving a classical cost of
$\bigO{n^3}$.  There is an interesting alternative classical algorithm
due to Floyd and Warshall that finds all the distances between all
pairs of vertices in a weighted graph.  Its cost is also $\bigO{n^3}$.


\head
3. Minimal Weight Spanning Trees
\endhead

Another common problem in classical graph theory is that of finding a
minimal weight spanning tree of a graph.  There is a very nice
classical algorithm for this due to Prim that goes as follows  (see \cite{G}):
$$\vbox{\vskip 10pt
\vbox{\algf}
\vskip 10pt\hskip 30pt
\hbox{\it Figure 4: Prim's algorithm.}
}$$
At the end of this algorithm, $F$ is a set of edges that comprise 
a minimal weight spanning tree of the original graph $(V,E)$, which
has been implicitly assumed to be a complete graph on $n=\card{V}$ 
vertices.  As in the case of Dijkstra's minimal weight path algorithm,
this algorithm consists of an search phase for each iteration and an
update phase for each iteration.  In the original algorithm of Prim,
just as in Dijkstra's original algorithm, the work for the search phase
and the work for the update phase are carefully balanced to be the same,
the total work being $\bigO{n^2}$.

There is a version of this algorithm that doesn't employ any updating
at the cost of having to search a larger space to find a minimum on
each iteration.  That algorithm goes as follows:
$$\vbox{\vskip 10pt
\vbox{\algg}
\vskip 10pt\hskip 30pt
\hbox{\it Figure 5: Prim's algorithm with no updating}
}$$ 

Essentially the same idea that works for a good quantum version of
Dijkstra's algorithm, as in the first section of this note, works to
give a good quantum version of Prim's algorithm.  In particular, the
following modification of Prim's algorithm is what seems to work:
$$\vbox{\vskip 10pt
\vbox{\algh}
\vskip 10pt\hskip 30pt
\hbox{\it Figure 6: Prim's algorithm with periodic updating}
}$$
The analysis of this algorithm is practically identical to that of
Dijkstra's algorithm, and again the parameter $k$ is determined to
equalize the work of searching for a minimum with the work of
updating.  If quantum minimum finding is used in steps (3a), (3b), and
(3e1a) instead of classical minimum finding, the optimal value of $k$
is again $\bigO{n^{1/2}}$ and the overall work for the quantum
algorithm in figure 6 here is $\bigO{n^{7/4}}$, which is an
improvement on the classical algorithm.

\head
4. Bipartite Graphs
\endhead

Another approach is to change the problem to one that is more
amenable to replacing the classical search with a quantum search.  The
idea is to balance the classical update cost per iteration with the
quantum search cost by introducing asymmetry into the problem in
Dijkstra's original algorithm.

In a bipartite graph, the set of vertices $V$ is divided into two
disjoint subsets $V_1$ and $V_2$, and all edges connect points in
$V_1$ with points in $V_2$ (and vice versa, as well in a directed
graph).  In a complete bipartite graph, each point in $V_1$ is
connected to every point in $V_2$, and conversely, each point in $V_2$
is connected to every point in $V_1$.  If $n_i=\card{V_i}$ for
$i=1,2$, then the number of edges in a complete bipartite graph is
$n_1n_2$, and in a complete bipartite digraph, the number of edges is
$2n_1n_2$.  If $n_1$ and $n_2$ are of very different sizes, then a
complete bipartite graph will be quite sparse.  In what follows, for
this section, the question of finding the lowest weight path in
complete bipartite graphs will be considered.  Since a complete
bipartite graph has edge set $E=(V_1 \times V_2) \cup (V_2 \times
V_1)$, it will be assumed that a pair of weight functions $\weight_1:V_1
\times V_2 \rightarrow \R^+$ and $\weight_2:V_2 \times V_1 \rightarrow \R^+$
are given.  The length of the shortest path $\lambda_i: V_i
\rightarrow \R^+$ from an initial vertex $v_0$ to elements in $V_i$ for
$i=1,2$ will be computed by the algorithm.

Assuming that $n_1<n_2$, the work for Dijkstra's algorithm is
$\bigO{n_1n_2\log n_2}$, with a sparse graph modification of the
algorithm using priority queues.  
There are certain cases in which this may not be the best classical
algorithm for finding the shortest path between two points.  The run
time for Dijkstra's algorithm may be optimal, but
the memory requirements may be overwhelming.
The following algorithm constructs the list of shortest distances
(and paths if a little additional information is kept) between a fixed
initial point $v_0\in V_1$ and all the other points in $V_1 \cup v_2$:
$$\vbox{\vskip 10pt
\vbox{\alge}
\vskip 10pt\hskip 25pt
\hbox{\it Figure 7: Dijkstra's algorithm for a bipartite graph with partial updating.}
}$$
The idea is that since $V_1$ is smaller than $V_2$, only the distances
in $V_1$ need to be updated.  Vertices in $V_2$ are viewed merely as
intermediate points along the way for paths that connect vertices in
$V_1$.

The analysis of this algorithm is only slightly more complicated than
the previous analyses.  Beginning with the idea that classical search
for the minimum of some function over a set of size $N$ requires
$\bigO{N}$ operations, it is clear that line (2a) requires work $\bigO{n_2}$
for each iteration of line (2), and since there are $n_1$ iterations
of line (2), the total work for the initialization of $\lambda(v)$ for
all $v\in V_1$, which is what goes on in lines (1), (2), (2a), and
(2b), is $\bigO{n_1n_2}$.  As for line (3), let $i=\card{S}$ on each
iteration.  Then for (3a), the search is over a set of size
$i\,(n_1-i)\,n_2$, which makes the total search cost for (3a) over all
the iterations in line (3) $\sum_{i=0}^{n_1}i\,(n_1-i)\,n_2
=\bigO{n_1^2n_2}$.  As for the update operations that start in
line (3), there is an subiteration over a set of size $n_1-i$ and
within each subiteration, there is a search over a set of size $n_2$.
Thus the total update cost is also
$\sum_{i=0}^{n_1}i\,(n_1-i)\,n_2=\bigO{n_1^2n_2}$.  Thus the
total (classical) algorithm cost is $\bigO{n_1^2n_2}$, but only
a memory of size $\bigO{n_1}$ is needed.  For the final stage (4) of the
algorithm that fills in the cost function for the elements of the second 
and larger part, $V_2$, the outer iteration is over a set of size $n_2$.
For each $u\in V_2$, the minimum in line (4a) is found over a set of 
size $n_1$, making the total (classical) work for this final phase 
of the algorithm $\bigO{n_1n_2}$, which is clearly not the dominant cost.

Now for the quantum costs.  The costs of initialization are $n_1$
iterations of search over a set of size $n_2$, so the quantum cost is
$\bigO{n_1n_2^{1/2}}$.  The search cost per iteration in line (3a)
is $\bigO{\sqrt{i\,(n_1-i)\,n_2}}$, making the total search cost
$$\sum_{i=0}^{n_1} \sqrt{i\,(n_1-i)\,n_2} 
\approx n_2^{1/2}\int_{0}^{n_1} \sqrt{x\,(n-x)}\, dx
= \bigO{n_1^2n_2^{1/2}}.$$
The update procedure, starting on line (3c) consists of subiterating 
over a set of size $n_1-i$, and for each subiteration, there is a search 
over a set of size $n_2$.  The total cost of updating is therefore 
$$\bigO{\sum_{i=0}^{n_1}\sum_{j=0}^i\sqrt{n_2}} = \bigO{n_1^2n_2^{1/2}}$$
which is the same as the search cost.  The total cost of this algorithm prior
to computing the minimum weight paths for all of $V_2$ is therefore
$\bigO{n_1^2n_2^{1/2}}$, with a memory cost of $\bigO{n_1}$.  
The final stage of the algorithm costs $n_2$ iterations of minimum finding 
over a set of size $n_1$, so the (quantum) work for filling in the cost function for
$V_2$ is $\bigO{n_2n_1^{1/2}}$.  The total cost of the whole algorithm is therefore
$\bigO{n_1^2n_2^{1/2}+n_2n_1^{1/2}}$.

How does this compare to the best classical cost, which is
$\bigO{n_1n_2}$ with a memory cost of $\bigO{n_1n_2}$, as well?  
Suppose that only the minimum weight paths to the other elements of 
$V_1$  are desired.  Then if
$\bigO{n_1^2n_2^{1/2}} < \bigO{n_1n_2}$, then the quantum algorithm wins.  This
occurs if $n_1 = \littleo{\sqrt{n_2}}$.  Thus for highly unbalanced
complete bipartite graphs, the quantum algorithm outperforms the
classical algorithm.

\head
Conclusion
\endhead

The quantum versions of Dijkstra's algorithm in section 1 and the
quantum version of Prim's algorithms in section 3 are for complete
graphs with a well defined weight function on all the edges.  The
improvements over the corresponding classical algorithms are also for
complete graphs.  If a graph is not complete, then the classical
minimal weight path algorithm and the classical minimal weight
spanning tree algorithms can effectively use the graph structure to
give better classical algorithms.  Essentially this is done through
updating only the vertices that are adjacent to the vertex being added
to the core set.  Classically, the use of priority queues lead to
these improvements.  It seems likely that these ideas can be
incorporated into future quantum algorithms for incomplete graphs,
but this is a topic for a future paper.  One particular case of
this for complete bipartite graphs was dealt with in section 4.

There are a number of closely related problems to finding minimal
weight paths in graphs.  Among these are the problems of deciding
graph connectivity and finding the shortest path between two points.
Determining the number of blocks in a graph follows directly from
being able to decide graph connectivity as does determining the set of
articulation points in a graph.  Presumably the quantum algorithms
given here can be readily extended to these problems, although the
details are yet to be worked out.

The whole field of quantum algorithmic graph theory has barely been
touched on here.  However, the algorithms developed in this note
should be regarded as strong evidence that many well known classical
graph theory algorithms have interesting quantum analogues.  Often
they will involve some nontrivial modifications of the classical
algorithm to make optimal use of the few tools currently available in
the quantum toolbox.  The optimistic view is that although the
modifications to the classical algorithms may be nontrivial, they
often are not excessively complicated either, as in the case of the
quantum versions of Dijkstra's algorithm and Prim's algorithm.

\enddocument

\end